\documentclass[preprint2]{aastex}

\begin{document}


\title{Detection of a very bright optical flare from the gamma-ray burst\\
   GRB 050904 at redshift 6.29\footnote{Partly based on observations made at the Observatoire de Haute Provence (France) 80cm telescope}}


\author{M. Bo\"er}
\affil{Observatoire de Haute-Provence (CNRS/OAMP),  F--04870 Saint Michel l'Observatoire, France}
\email{Michel.Boer@oamp.fr}

\author{J.L. Atteia}
\affil{LATT, CNRS-Observatoire Midi-Pyr\'en\'ees, Universit\'e Paul Sabatier, 14 Av. Edouard Belin, 31400 Toulouse, France}
\email{atteia@ast.obs-mip.fr}

\author{Y. Damerdji\altaffilmark{2}}
\affil{Observatoire de Haute-Provence (CNRS/OAMP)  F--04870 Saint Michel l'Observatoire, France}
\email{yassine.damerdji@oamp.fr}

\author{B. Gendre}
\affil{Istituto di Astrofisica Spaziale e Fisica Cosmica, sede di Roma, INAF, via fosso del cavaliere 100, 00133 Roma, Italy}
\email{Bruce.Gendre@rm.iasf.cnr.it}

\author{A. Klotz\altaffilmark{2}}
\affil{Observatoire de Haute-Provence (CNRS/OAMP),  F--04870 Saint Michel l'Observatoire, France}
\email{alain.klotz@free.fr}

\and

\author{G. Stratta}
\affil{LATT, CNRS-Observatoire Midi-Pyr\'en\'ees, Universit\'e Paul Sabatier, 14 Av. Edouard Belin, 31400 Toulouse, France}
\email{gstratta@ast.obs-mip.fr}









\altaffiltext{2}{Centre d'Etude Spatiale des Rayonnements (CNRS/UPS), Observatoire Midi-Pyr\'en\'ees, Universit\'e Paul Sabatier, BP 4346, F--31028 -Toulouse Cedex 04, France}


\begin{abstract}

In this letter we discuss the flux and the behavior of the bright optical flare emission detected by the 25 cm TAROT robotic telescope during the prompt high-energy emission and the early afterglow.  We combine our data with simultaneous observations performed in X-rays and we analyze the broad-band spectrum.  These observations lead us to emphasize the similarity of GRB 050904 with GRB 990123, a remarkable gamma-ray burst whose optical emission reached 9th magnitude. While GRB 990123 was, until now, considered as a unique event, this observation suggests the existence of a population of GRBs which have very large isotropic equivalent energies and extremely bright optical counterparts. The luminosity of these GRBs is such that they are easily detectable through the entire universe. Since we can detect them to very high redshift even with small aperture telescopes like TAROT, they will constitute powerful tools for the exploration of the high-redshift Universe and might be used to probe the first generation of stars.

\end{abstract}


\keywords{Gamma Ray: bursts}



\section{Introduction}

Gamma-Ray Bursts (hereafter GRBs) are powerful stellar explosions which are easily detectable as they are accompanied by a burst of high-energy photons usually lasting several seconds. The high energy emission occurs within a jet of ultra-relativistic material ejected towards the Earth by the exploding star.
Late afterglow emission is explained as the forward shock component of external
shock produced by the interaction of the expanding fireball with the external medium.
Very early optical emission is thought to be an effect of the reverse shock component of the external shock. Alternatively it can be explained as the low energy tail of the high energy emission of the internal shock \citep{mes99,katz94}.

The event of September 4th, 2005 (\object{GRB 050904}) was detected by the SWIFT/BAT experiment at 01:51:44 UT \citep{cusu05}. The alert was distributed via the Gamma-ray burst Coordinate Network (GCN) at 01:53:05 UT.
Five seconds later, TAROT (Telescope Action Rapide pour les Objets Transitoires - Rapid Action Telescope for Transient Objects \citep{boer99}), a 25cm robotic telescope located on the plateau du Calern in Southern France started to observe the field. In the meantime SWIFT slewed to the source direction.
Observations with its narrow field instruments (XRT, the X-Ray Telescope, and UVOT, the UV/Optical Telescope) started about 160 seconds after the burst triggered the instrument.
BOOTES-1B 30 cm robotic telescope in Southern Spain observed simultaneously to TAROT with the R-band filter although no detection was achieved \citep{jel05}.
Large telescopes started to image the field three hours later and discovered a bright, fading near-infrared source \citep{cusu05,tagl05}. The lack of counterpart at visible wavelengths was quickly attributed to the Lyman alpha absorption in a highly redshifted source \citep{hais05a}. The Subaru telescope measured the spectroscopic redshift of the source \citep{kawa05} that was found \cite{hais05b} to be at a z = 6.29, corresponding to an age of the Universe which is only 7$\%$ of the present epoch. To date GRB 050904 is the gamma-ray burst with the highest measured redshift and the farthest cosmological source ever observed with a small (25cm) telescope.
In this letter we present the observations made by TAROT (section 2). The results of a joint temporal and spectral analysis of both the visible and SWIFT-XRT data are presented in section 3. We then discuss these results in the framework of theoretical models, and we show evidences that both GRB 990123 and GRB 050904 might belong to an overluminous class of bursts.

 \section{Observations and analysis}


TAROT acquired data from 86 to 1666 seconds after the trigger. Twenty-two unfiltered images were recorded with exposure times varying from 15 to 90 sec (as time passes, the exposure time was increased to optimize the sensitivity). The first images were acquired while the burst was still active at gamma-ray energies. In order to increase the signal to noise ratio we co-added these images and computed a series of six frames which are called T1-T6 hereafter.  Table 1 gives the main characteristics of the data. The optical transient is detected in images T2, T3 and T4, i.e. from 150 to 589 sec after the trigger with equivalent I-band magnitude (see below) varying from $>$15.2 to 14.1 mag. Figure 1 displays the optical transient as seen in an 828 second long co-added frame (T2-T5). Because of the high redshift of the source and of the quantum yield of the CCD camera, the recorded photons lie all in the range from 880 to 1000 nm. This explains why the BOOTES telescope, which is slightly larger (30cm diameter) than TAROT (25cm), and started observations 124s after the burst onset, did not detect it \citep{jel05}: the use of an R-band filter makes the detection of such a high redshifted object almost impossible, because of the Lyman alpha cutoff. In order to compute magnitudes relative to a standard system we used the 80 cm telescope of the Observatoire de Haute-Provence during the night of September 12th, 2005, to measure the VRI magnitudes and the spectrum of three nearby stars. Then we deduced the spectral type of these stars and we modeled the flux ratio of the GRB source relative to the stars for the unfiltered TAROT images. We assumed a spectral shape $f_{\nu}\propto\nu^{-0.7}$ for the GRB spectrum. We checked that the flux values are rather insensitive to the spectral index of the GRB: uncertainties of $\pm$0.5 on the spectral index induce errors lower than 20$\%$ on the estimated flux at 9500\AA, thanks to the narrow range of wavelength used by the detector to measure the flux received from the source.

The observations reported here provide the only optical detections available for GRB 050904 during the first minutes after the trigger; they offer a continuous coverage of the prompt emission, of the reverse shock and of the early afterglow.
In order to obtain the broad-band spectrum of the burst, we also analyzed X-ray data collected by the SWIFT XRT instrument (see also \citet{wats05}). SWIFT XRT data were extracted using the XRT Tools version 2.1 and standard filtering criteria\footnote{see http://swift.gsfc.nasa.gov/docs/swift/analysis}. The bin sizes were chosen in order to contain 20 source counts each. We ignored bins below 0.3 keV (Window Timing mode) and 0.5 keV (Photon Counting mode) during the fit. Count-rate light-curves (mode dependant) were converted into a single flux light-curve (mode independent) by using the conversion factors derived from our spectral analysis. Spectra were fitted with a power law model taking into account the absorption both from the Galaxy \citep{dick90} ($N_H = 4.97\times10^{20}$ cm$^{-2}$) and the host.
The temporal evolution of the optical emission is displayed in figure 2, along with the X-ray light-curve.

\section{Results}

The optical emission can be described as a gradual increase of the optical emission during the first 150 seconds of the prompt emission (T1) followed by a plateau lasting 250-300 sec (T2-T3); a bright flare takes place before the end of the prompt gamma-ray emission\footnote{In this analysis we follow Cusumano et al. (2005) considering that the prompt emission ends after the bright peak which culminates 467 sec after the trigger (see also Zou et al. 2005} (T4). We marginally detect the optical transient after the end of the prompt emission with a signal to noise ratio S/N = 3.2 for 380 sec (T5). This emission is not detected in the subsequent time interval lasting 680 sec (T6). Taking the trigger instant as the origin of times, the rapid decrease between T4 and T5 implies a decay index greater than 3. We computed the flux expected from the backward extrapolation of the afterglow measured from 3 to 12 hours after the burst \citep{hais05b,tagl05} (dotted line): the peak of the actual optical emission exceeds the flux computed from the afterglow extrapolation by a factor of five at least. This should be considered as a lower limit since the interval containing the peak of the optical emission is integrated over 140 sec.

\section{Discussion and conclusion}

Four gamma-ray bursts have been observed at optical wavelengths while the high-energy source was still active: GRB 990123 at z=1.60 \citep{aker99}, GRB 041219a at an unknown redshift \citep{vers04}, GRB 050401 at z=2.90 \citep{ryko05a} and finally GRB 050904 at z=6.29 which we present here. All these events exhibit variable optical emission on short time scales. The flux density at the maximum of the optical emission are respectively 1080 mJy (V band), 10 mJy (R band), 0.6 mJy (R band), and 48 mJy (I band). These numbers translate into 1080, 3.1, 2.6, and 1300 mJy if we scale the four gamma-ray bursts at z=1.60, the redshift of GRB 990123; here we assume that GRB 041219a has a redshift of z=1 and we neglect any k-correction. Taken at face, these numbers show that the optical flares associated with GRB 050904 and GRB 990123 were very bright. In both events the optical flare is detected near the end of the prompt emission (right after a high energy peak) and exceeds significantly the backward extrapolation of the late-time afterglow. At redshift z=1.60, the magnitude of the flare detected by TAROT would have been comparable with the one of the optical transient associated with GRB 990123. Other similarities between these two bursts include a high rest-frame peak energy (2000 keV for GRB 990123, and $>$1500 keV for GRB 050904), and a large isotropic-equivalent energy release ($2.8\times10^{54}$ erg for GRB 990123, and $\sim 10^{54}$ erg for GRB 050904). A major difference between both events may be the plateau observed for GRB 050904 during the gamma-ray event (T2-T3), both at X-ray and optical wavelengths. This feature is not apparent on the ROTSE data from GRB 990123. However this is easily understandable since the TAROT dead time between two consecutive exposures was only 5 seconds while it was 20 seconds for ROTSE. This makes the absence of the plateau in GRB 990123 inconclusive.

The simultaneous observations of SWIFT and TAROT allow the measure of the broad-band spectrum of GRB 050904 and its evolution. Figure 3 displays the $\nu f_{\nu}$ spectra computed for intervals T2 to T6 of Table 1. We can derive several features from these spectra: {\it i)} during interval T2 the slope of the X-ray spectrum ($\alpha=$-1.19$\pm$0.09) is compatible with the spectral slope measured in gamma-rays with BAT (-1.22$\pm$0.1); {\it ii)} the  substantial hydrogen column density, in excess of the galactic value, apparent during intervals T2 and T3 disappears afterward ; finally, {\it iii)} the optical emission stands well above the extrapolation of the X-ray spectrum, even if we deconvolve for the absorption. This is clearly illustrated by the slope of the broad-band $\nu f_{\nu}$ spectrum (between 950 nm and 1 keV), which is +0.27$\pm$0.05; +0.22$\pm$0.05; -0.33$\pm$0.05 during T2, T3, and T4, respectively (Figure 3). This last feature was also observed in GRB 990123
\citep{cors05}.

The optical emission of GRB 990123 has been explained by reverse shock emission \citep{sari99}. Wei et al. (2005) suggested that also for GRB 050904 the optical emission can be explained with the reverse shock model. In this case, the reverse shock would have started between intervals T1 and T2, about 150 sec after the trigger (20 seconds in the rest-frame). A long lasting optical emission could be produced when the reverse shock propagates into the various layers of the ejecta \citep{naka05}.
The reverse shock could also explain the fast decrease of the afterglow emission after the optical flare.
However, in the reverse shock interpretation of the optical emission, we have to admit that the temporal coincidence of the optical flash with an X-ray peak, which is part of the prompt emission, is fortuitous
\citep{wei05}.

Alternatively, this coincidence could find a natural explanation if the optical emission is attributed to the prompt emission \citep{vers04}. In particular, Wei et al. (2005) proposed as an alternative model to the reverse shock, where the optical flare is produced by a late internal shock from a re-activity of the internal engine and the X-ray flare is  produced by the synchrotron-self-Compton mechanism (see also Zou et al. 2005).

These observations demonstrate that continuous joint optical and X-ray coverage with a time resolution of several seconds is essential to understand the origin of the GRB prompt emission and of the early afterglow. Given the intrinsic brightness of GRB 050904, the measure of accurate optical light curves with a temporal resolution of a few seconds was within the reach of 1 meter class robotic telescopes working in the optical and in the near infrared. If GRB 050904 were at redshift unity, TAROT would have produced detailed time-resolved observations of the prompt emission.
The lack of bright optical transients, similar to those of GRB 990123 and GRB 050904, for the majority of GRBs at redshifts 1 or 2, implies that optically bright GRBs (hereafter OB-GRBs) are uncommon \citep{romi05}.
Since the origin of the prompt optical emission has not yet a univocal interpretation, it seems difficult to identify the factor(s) responsible of their brightness. We note however that GRB 990123 and GRB 050904 have in common large isotropic equivalent energies and very high intrinsic peak energies (about 2 MeV). It is also interesting to notice that while at very different distances, the optical behavior of both events is remarkably similar. Two events are not sufficient to tell whether the bright optical emission is always associated with very high peak energies or whether we must call upon other factors, like a different environment and/or distinct progenitors. However, the TAROT observations show that GRB 990123 was not a once-in-a-lifetime event \citep{ryko05b}. The existence of GRBs with bright optical afterglows has many important consequences: {\it i)} bright optical flares can be monitored with temporal resolutions of the order of the second, a sampling which is essential to understand the origin of the prompt optical emission; {\it ii)} the detection of OB-GRBs at high redshift may be the best way to hold the promises of GRBs for cosmology, since the measure of GRB redshifts beyond redshift 5 requires intrinsically bright optical and infrared emission; {\it iii)} finally, the importance of OB-GRBs is confirmed by the fact that GRB 990123 was the first GRB for which the prompt optical emission could be detected, while GRB 050904 was the first GRB at redshift $>$5 whose redshift could be measured and whose light curve could be studied in details. Thanks to the combination of high-energy space experiments and of the expected increasing coverage of fast robotic telescopes devoted to the follow-up at optical and near infrared wavelengths we expect that more breakthroughs will result from the detection of new optically bright GRBs in the future.



\acknowledgments
BG and GS acknowledge the support by the EU Research and Training Network "Gamma-Ray Bursts, an enigma and a tool". TAROT has been funded by the Centre National de la Recherche Scientifique, Institut National des Sciences de l'Univers (INSU-CNRS), and built thanks to the support of the Technical Division of INSU-CNRS.
Correspondence and requests for data should be addressed to MB.



{\it Facilities:} \facility{OHP:0.8m ()}, \facility{Swift ()} , \facility{TAROT ()}




\clearpage



\begin{figure}
\epsscale{.80}
\plotone{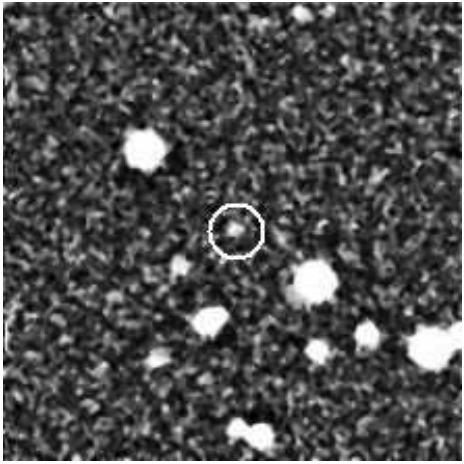}
\caption{Image of the field of GRB 050904 from TAROT. We co-added all frames from 150 to 978 seconds after the Swift trigger. This interval covers the prompt gamma-ray emission as well as the X-ray flash event. The size of the image is 6 x 6 arcmin centred on RA = 00 h 54 min 50.9 s, DEC = +14$^{\circ}$ 05$'$ 09$"$ (J2000.0). North is up, east is left. The spatial sampling is 3.3 arcsec/pixel. The afterglow is visible within the white circle.\label{fig1}}
\end{figure}

\clearpage


\begin{figure}
\plotone{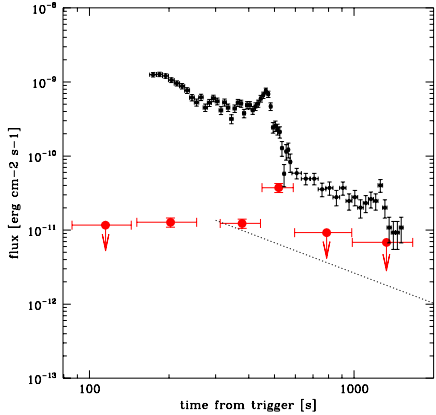}
\caption{Optical (I-band, red) and X-ray (0.5-10 keV, black) light curves of GRB 050904 showing the end of the prompt emission and the transition to the afterglow. The optical emission was measured with TAROT from 86 seconds to 28 minutes after trigger. The X-ray light curve (0.5-10 keV, black) was measured with the XRT from 160s to 33 minutes after the trigger (2). The dotted line shows the extrapolation of the late afterglow (measured 3 hours after the trigger), assuming a power law temporal decay with an index of -1.36 \citep{hais05b}. \label{fig2}}
\end{figure}

\begin{figure}
\epsscale{.80}
\plotone{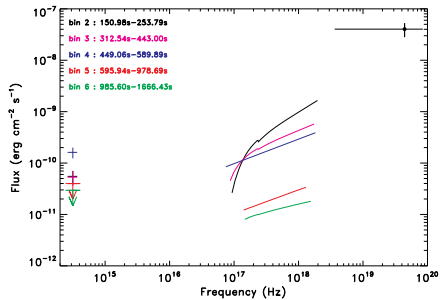}
\caption{Broadband spectrum of GRB 050904 in the $\nu f_{\nu}$ space, corrected for the galactic absorption. We display optical (I band) data taken by TAROT and X-ray (0.3-10.0 keV) data taken by the SWIFT X-Ray Telescope during temporal bins T2 to T6 described in Table 1. We have also plotted the gamma-ray observation made by BAT on-board SWIFT during the second TAROT temporal bin (T2). For clarity, only the best fit models are presented for X-ray data. \label{fig3}}
\end{figure}

\clearpage

\begin{table}
\begin{center}
\caption{Journal table of the TAROT observations together with the main characteristics of the co-added images. Columns 1, 2 and 3 indicate respectively the interval designation, the beginning and the end of co-added exposures since the GRB trigger. Column 4 gives the flux density (1$\sigma$ errors)of the optical transient at 9500\AA  as deduced from the spectrophotometric calibration of three nearby stars (see text). Column 5 gives the equivalent I-band magnitudes (see text) \label{tbl-1}}
\begin{tabular}{crrrrr}
\tableline\tableline
Interval & Start time & End time & Flux density at 9500\AA & I-band \\
    &  [s]      & [s]   & [$10^{-15}$erg/cm$^2$s\AA]  & [mag] \\
\tableline
T1 & 86   & 144  &   $\la$5.3          &    $> $15.3    \\
T2 & 150  & 253  &   $5.8^{+1.6}_{-1.2}$   &    15.22$\pm$ 0.26\\
T3 & 312  & 443  &   $5.6^{+1.8}_{-1.4}$   &    15.26$\pm$ 0.30\\
T4 & 449  & 589  &   $17\pm4$          &    14.07$\pm$ 0.24\\
T5 & 595  & 978  &   $\la$4.2          &    $> $15.5    \\
T6 & 985  & 1666 &   $\la$3.1          &    $> $15.8    \\
\tableline
\end{tabular}
\end{center}
\end{table}


\clearpage

\begin{deluxetable}{ccrrrrrr}
\tabletypesize{\scriptsize}
\tablecaption{Results from the X-ray spectral analysis. Column 1 indicates the time interval taken for the TAROT data (see table 1), column 2 the corresponding interval for the SWIFT data analysis, column 3 the extragalactic absorption (at a redshift of z=6.29) in excess to the galactic absorption (fixed at 4.97 $\times10^{20}$ cm$^{-2}$, \citet{dick90}), column 4 the energy spectral index and column 5 the reduced chi-square for the fit. Column 6 to 8 give the flux computed respectively in the spectral band 0.5-2.0 keV, 2.0-5.0 keV, 5.0-10.0 keV. The data quality and/or calibration status did not allow us to derive the extragalactic absorption for intervals T5 and T6.\label{tbl-2}}
\tablewidth{0pt}
\tablehead{
\colhead{Interval} & \colhead{$t-t_0$} & \colhead{$N_H$} & \colhead{$\alpha$} & \colhead{$\chi_{\nu}^2(\nu)$} &
\colhead{Flux(0.5-2.0 keV)} & \colhead{Flux(2.0-5.0 keV)} & \colhead{Flux(5.0-10.0 keV)}
}
\startdata
T2 & 169.0-253.8  & $7.3^{+4.9}_{-3.9}$ & $0.19\pm0.09$ & 1.04(136) & $39.9\pm0.7$ & $74\pm2$ & $96\pm2$   \\

T3 & 312.5-443.0  & $2.8^{+2.9}_{-2.2}$ & $0.44\pm0.08$ & 1.27(120) & $26.5\pm0.5$ & $34.5\pm0.7$ & $40.7\pm0.8$   \\

T4 & 449.1-582.0  & $<0.37$ & $0.52\pm0.06$ & 1.07(92) & $20.8\pm0.5$ & $23.7\pm0.6$ & $26.4\pm0.6$   \\

T5 & 582.0-978.7  & - & $0.53\pm0.09$ & 1.4(12) & $2.2\pm0.2$ & $2.5\pm0.2$ & $2.7\pm0.2$   \\

T6 & 985.6-1666.4  & - & $0.7\pm0.2$ & 0.79(17) & $1.39\pm0.06$ & $1.35\pm0.06$ & $1.35\pm0.06$   \\
\enddata
\end{deluxetable}

\end{document}